\documentclass{llncs}
\usepackage{epsf,epsfig}
\title{The case of the missing neutrality}
\author{Russell K. Standish}
\institute{School of Mathematics\\
University of New South Wales,Sydney, 2052,Australia\\
\email{R.Standish@unsw.edu.au}\\http://parallel.hpc.unsw.edu.au/rks}

\begin{document}
\maketitle
\begin{abstract}
  The concept of neutral evolutionary networks being a significant
  factor in evolutionary dynamics was first proposed by Huynen {\em et
    al.} about 7 years ago. In one sense, the principle is easy to
  state --- because most mutations to an organism are deleterious, one
  would expect that neutral mutations that don't affect the phenotype
  will have disproportionately greater representation amongst successor
  organisms than one would expect if each mutation was equally likely.

  So it was with great surprise that I noted neutral mutations being
  very rare in a visualisation of phylogenetic trees generated in {\em
  Tierra}, since I already knew that there was a significant amount of
  neutrality in the Tierra genotype-phenotype map. The paper reports
  on an investigation into this mystery.
\end{abstract}

\section{Introduction}

The influence of {\em neutral networks} in evolutionary processes was
first elucidated by Peter Schuster's group in Vienna in
1996\cite{Huynen-etal96,Reidys-etal96}. Put simply, two {\em
  genotypes} are considered {\em neutrally equivalent} if they map to
the same {\em phenotype}. A {\em neutral network} is a set of
genotypes connected by this neutrality relationship on links with
Hamming distance 1 (ie each link of the network corresponds to a
mutation at a single site of the genome).

These researchers noted that evolution tended to proceed by diffusion
along these neutral networks, punctuated occasionally by rapid changes
to phenotypes as an adaptive feature is discovered. The similarity of
these dynamics with the theory of Punctuated
Equilibria\cite{Eldridge85} was noted by Barnett\cite{Barnett98}. It
was also noted that if a {\em giant network} existed that came within
a hop or two of every possible genotype, evolution will be
particularly efficient at discovering solutions.

Most work on neutrality in evolution uses the genotype-phenotype
mapping defined by folding of RNA\cite{Schuster-etal94}. This mapping
is implemented in the open source Vienna RNA
package\footnote{http://www.tbi.univie.ac.at/\~{}ivo/RNA}, so is a
convenient and well known testbed for ideas of neutrality in evolution.

Also in 1996, I developed a definition of the genotype-phenotype
mapping for Tierra, which was first published in
1997\cite{Standish97b}. I noticed the strong presence of neutrality in
this mapping at that time, which was later exploited to develop a
measure of complexity of the Tierran
organism\cite{Standish99a,Standish03a}. In 2002, I started a programme
to visualise Tierra's phylogenetic trees and neutral
networks\cite{Standish02c} in order to ``discover the unexpected''.
Two key findings came out of this: the first being that Tierra's
genebanker data did not provide clean phylogenetic trees, but had
loops, and consisted of many discontinuous pieces. This later turned
out to be due to Tierra's habit of reusing genotype labels if those
genotypes were not saved in the genebanker database. This might happen
if the population count of that genotype failed to cross a threshold.
This is all very well, except that a reference to that genotype exists
in the parent field of successor genotypes. The second big surprise
was the paucity of neutral mutations in the phylogenetic tree. We
expect most mutations to an organism to be deleterious, and so expect
that neutral mutations will have disproportionately greater
representation amongst successor genotypes than one would expect if each
mutation was equally likely.

\section{Neutrality in Tierra}

{\em Tierra}\cite{Ray91} is a well known artificial life system in which
small self-replicating computer programs are executed in specially
constructed simulator. These computer programs (called digital
organisms, or sometimes ``critters'') undergo mutation, and radically
novel behaviour is discovered, such as {\em parasitism} and {\em
  hyperparasitism}.

It is clear what the genotype is in Tierra, it is just the listing of
the program code of the organism. The phenotype is a more diffuse
thing, however. It is the resultant effect of running the computer
program, in all possible environments. Christoph Adami defined this
notion of phenotype for a similar artificial life system called {\em
  Avida}\cite{Adami98a}. In Avida, things are particularly simple, in
that organisms either reproduce themselves at a fixed replication rate,
or don't as the case may be, and optionally perform range of
arithmetic operations on special registers (defined by the experimenter).

In Tierra, organisms do interact with each other via a template
matching mechanism. For example, with a branching instruction like
\verb+jmpo+, if there is a sequence of \verb+nop0+ and \verb+nop1+
instructions (which are nooperations) following the branch, this
sequence of 1s and 0s is used as a template for determining where to
branch to. In this case the CPU will search outwards through memory
for a complementary sequence of \verb+nop0+s and \verb+nop1+s. If the
nearest complementary sequence happens to lie in the code of a different
organism, the organisms interact.

To precisely determine the phenotype of a Tierran organism, one would
need to execute the soup containing the organism and all possible
combinations of other genotypes. Whilst this is a finite task, it is
clearly astronomically difficult. One means of approximation is to
consider just interaction of pairs of genotypes (called a tournament).
Most Tierran organisms interact pairwise --- very few triple or higher
order interactions exist. Similarly, rather than running tournaments
with all possible genotypes, we can approximate matters by using the
genotypes stored in a genebanker database after a Tierra
run. In practice, it turns out that various measures, such as the
number of neutral neighbours, or the total complexity of an organism
are fairly robust with respect to the exact set of organism used for
the tournaments.

So the procedure is to pit pairwise all organisms in the genebanker
against themselves, and record the outcome in a table (there is a
small number of possible outcomes, which is detailed in
\cite{Standish97b}). A row of this table is a phenotypic signature for
the genotype labeling that row. We can then eliminate those genotypes
with identical signatures in favour of one canonical genotype. This
list of unique phenotypes can be used to define pragmatically a test
for neutrality of two different organisms. Pit each organism against
the list of unique phenotypes, and if the signatures match, we have
neutrality.  The source code for this experiment is available from the
author's
website.\footnote{http://parallel.hpc.unsw.edu.au/getaegisdist.cgi/getsource/eco-tierra.3,
  version 3.D3}

Tierra has three different modes of mutation:
\begin{description}
\item[Cosmic Ray] A site of the soup is randomly chosen and mutated;
\item[Copy] Data is mutated during the copy operation;
\item[Flaw] Instructions occasionally produce erroneous results
\end{description}
Furthermore, in the case of cosmic ray and copy mutations, a certain
proportion of mutations involve bitflips, rather than opcodes being
substituted uniformly. This proportion is set as a parameter in the
soup\_in file (\verb+MutBitProp+) --- in these experiments, this
parameter is set to zero.

In order to study the issue of whether neutrality is greater or less
than expected in Tierra, I generated three datasets with each of the 3
modes of mutation operating in isolation. The sizes of each data set
was 69,139, 87,003 and 198,982 genotypes respectively, 
generated over a time period of about 1000 million executed instructions.
Genebanker's threshold was set to zero, so all genotypes were
captured. This led to a proper phylogenetic tree. After performing a
neutrality analysis, a set of 83, 86 and 158 unique phenotypes was extracted as
the test set for the tournaments.

\begin{figure}
\centerline{\epsfbox{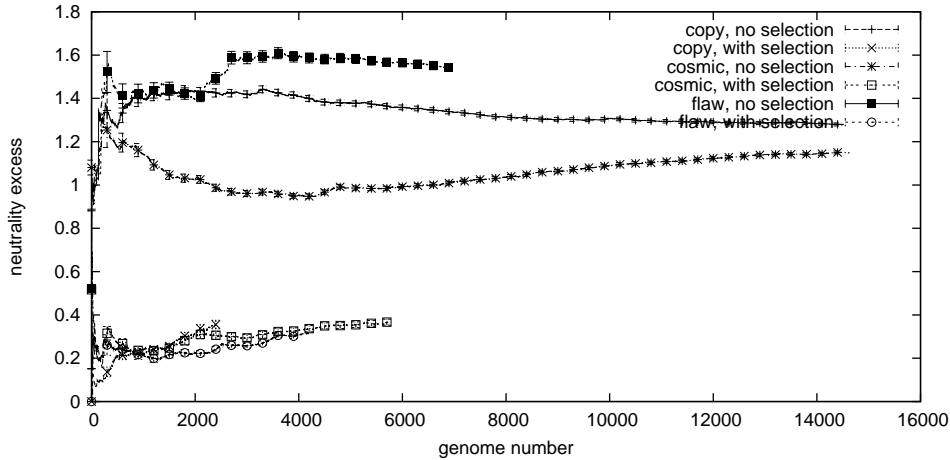}}
\caption{Running average of neutrality excess ($\langle
  r_i\rangle$). Genomes are ordered according to size, and neutrality
  excess is averaged over all genomes to the left of that data
  point. Three different datasets are analysed, with each of the three
  modes of mutation turned on. Then the datasets are further filtered
  to only include offspring whose maximum population count is greater
  than 1, ie selection is operating.
}
\label{nn-ratio}
\end{figure}

Since the neighbourhood size increases exponentially with
neighbourhood diameter, I restrict analysis to single site, or point mutations.
In each data set, around 7\% of these genotypes were
created  were created by a mutation at a single
site and were neutrally equivalent to its parent. 
 For each of these, I compute the number of neutral
neighbours $n_i$ existing in the 1 hop neighbourhood of the parent
genotype $i$, which is of size $32^{\ell_i}$, where $\ell_i$ is
the length of the genome. For a given parent $i$, the ratio 
\begin{equation}\label{neutrality excess}
r_i=\frac{\nu_i32^{\ell_i}}{o_in_i}
\end{equation}
gives the proportion of neutral links actually followed relative to
the number of neutral links available ({\em neutrality excess}), where
$\nu_i$ is the number of neutrally equivalent offspring, and $o_i$ the
total number of offspring and the size of the 1 hop neutral
neighbourhood. Fig.~\ref{nn-ratio} shows the running average of this
quantity over these transitions, with the genotypes numbered in size order.

In this analysis, no selection is operating, so one would expect that
the neutrality excess should be identical to 1. However, in the case
of instruction flaws, it is rather unpredicatable what the effect
is. In the case of cosmic ray mutations, 50\% of time one would expect
the parent to be mutated, rather than the daughter. In the case of a
mutation affecting a crucial gene of a parent genotype, the organism
may not be able to reproduce at all, thus favouring neutral
mutations. Only copy mutations should affect all sites of the genome
equally, leading to a neutrality excess equal to one. The measured
value, however is about 1.3, substantially greater than one. The
reason for this is not known at this point in time.

The the datasets were further subsetted to include just those
transitions whose daughter organism successfully reproduced, ie with a
maximum population count greater than 1. The neutrality excess in this
case is substantially less than 1, so something in Tierran evolution
is favouring nonneutral evolution.

\section{Vienna RNA folding experiments}

It is quite well known that evolution using the RNA folding
map\cite{Schuster-etal94} exhibits a great deal of neutrality, at
least for a standard genetic algorithm optimising a well defined
fitness function. Tierra is a coevolutionary system, and does not have
a well defined fitness function. Rather, the chance of an organism
surviving at each time step depends on what other organisms are in the
environment at the time, and has a significant component of
contingency.

One possible cause for the repression of neutrality in Tierra is this
coevolutionary nature. The reasoning follows from the idea of the {\em
  Red Queen effect}\cite{vanValen73}. This says that organisms must
continuously evolve just to remain adaptive.\footnote{Like the Red Queen in
Lewis Caroll's {\em Through the Looking Glass}, who had to keep
running, just to stay where she was.} In such a circumstance, neutral
evolution is maladaptive, and surely be suppressed. 

A convincing argument in favour of this hypothesis would be the
demonstration of a coevolutionary system based on the RNA folding map
that exhibited this repression of neutrality. Whilst I haven't
achieved this goal, I will report on a couple of attempts. 

The first attempt is an instantiation of the simplest possible
coevolutionary system. It consists of two populations: a tracker
population $T$ which attempts to be as similar to the other population
as possible, and an evader population $E$ that attempts to be as
different from the tracker population as possible. The Vienna RNA
folding library is used, and fitness functions $f_T(x,E)$, $f_E(x,T)$
defined for the trackers and evaders respectively, based on the
average distance between their phenotypes:
\begin{eqnarray}
  f_T(x,E) &=& -\frac1{N_E}\sum_{y\in E} d(x,y),\, \forall x\in T \nonumber\\
  f_E(x,T) &=& \frac1{N_T}\sum_{y\in T} d(x,y),\, \forall x\in E
\end{eqnarray}
where $N_T$ and $N_E$ are the population counts of trackers and
evaders respectively, and $d(x,y)$ is the {\em string edit
  distance}\footnote{Please consult the Vienna RNA package
  documentation for a precise definition of string edit distance}
between the folded structure of $x$ and $y$.

\begin{figure}
\centerline{\epsfbox{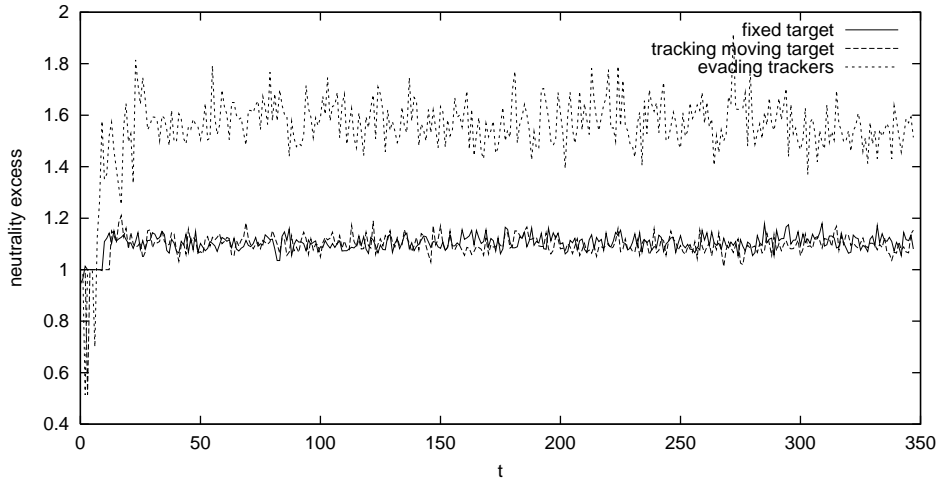}}
\caption{Neutrality excess for the RNA folding experiment, as a
  function of time. Separate lines are plotted for trackers, evaders
  and trackers tracking a fixed target.}
\label{rna}
\end{figure}

I implemented a simple genetic algorithm with just a point mutation
operator. All RNA strings have the same length. During reproduction,
each RNA string copies itself, possibly with a mutation. During the
selection step, the least fit 50\% of organisms are culled, bringing
the population count back to the starting value. For the results
presented in Fig. \ref{rna}, the GA parameters are shown in Table
\ref{GA parm}. Source code for this experiment is available from the
author's website.\footnote{http://parallel.hpc.unsw.edu.au/getaegisdist.cgi/getsource/rnafold/,
  version D1}

\begin{table}
\centerline{
\begin{tabular}{lr}
\hline
String length: & 20\\
Population size: & 100\\
Mutation probability per site: & 0.1 \\
\hline
\end{tabular}
}
\caption{Genetic Algorithm parameters for the RNA folding experiment
  reported in the paper}
\label{GA parm}
\end{table}

Fig. \ref{rna} shows the neutrality excess for trackers and evaders,
defined in the same way as eq (\ref{neutrality excess}). A third
baseline run of a single population tracking a fixed target is also
shown (corresponding to a classic fixed fitness function genetic
algorithm). The most obvious thing about these results is that
neutrality is highly adaptive to evaders, who clearly are trying to
make themselves occupy as small a footprint in phenotype space as
possible. Given the evaders propensity to stay still, trackers will
tend to behave like their fixed target counterparts.

This indicated that trackers were dominating over the evaders. 
Another experiment I performed was where all coevolving populations
were symmetric. In this case, each population would track a second and
evade a third population. The simplest arrangement of these has 3
coevolving populations in a ``Rock, Scissors, Paper'' configuration, so no one
population dominates. I also tried other combinations up to 10
separately coevolving populations, wired up randomly to each other
(but fixed at the start of the experiment). In all cases, the results
were much the same --- the neutrality excess was less than for the
trackers in Fig. \ref{rna}, but still just slightly greater
than 1.

So how do can we introduce the Red Queen effect to this system? One
possibility I haven't explored as yet is to somehow give evaders room
to move, perhaps by including a genome lengthening operator as part of
the GA.

\section{Conclusion}

The suppression of neutrality in Tierran evolution is a real
effect. It is quite likely that this is a Red Queen effect, with
organisms needing to change to remain adaptive. Experiments with using
the RNA folding map to try to reproduce this effect have proven
inconclusive. However, it was noted that there is significant evolutionary
pressure to increase neutrality in evading populations.

\section*{Acknowledgments}

I would like to thank the {\em Australian Centre for Advanced
  Computing and Communications} for a grant of computing time used in
  this project.

\bibliographystyle{plain}
\bibliography{rus}

\end{document}